\documentclass[aps,prd,superscriptaddress,nofootinbib,amsmath,amsfonts,preprintnumbers,groupedaddress,showpacs,9pt,english]{revtex4-1}
\usepackage{amsmath}
\usepackage{amssymb}
\usepackage{booktabs}
\usepackage{babel}
\usepackage{wrapfig}
\usepackage{cancel}

\usepackage{relsize,exscale}
\makeatletter
\newcommand{\beq}{\begin{equation}}
\newcommand{\eeq}{\end{equation}}
\newcommand{\bea}{\begin{eqnarray}}
\newcommand{\eea}{\end{eqnarray}}

\newcommand{\comm}[1]{}

\usepackage{array,multirow,graphicx}
\usepackage{dcolumn}
\usepackage{newlfont}
\usepackage{bm}
\usepackage[colorlinks,citecolor=blue,urlcolor=blue,linkcolor=blue]{hyperref}
\usepackage[figtopcap]{subfigure}
\usepackage{color}

\usepackage{scalerel}
\usepackage{tikz}
\usetikzlibrary{svg.path}
\definecolor{orcidlogocol}{HTML}{A6CE39}
\tikzset{
  orcidlogo/.pic={
    \fill[orcidlogocol] svg{M256,128c0,70.7-57.3,128-128,128C57.3,256,0,198.7,0,128C0,57.3,57.3,0,128,0C198.7,0,256,57.3,256,128z};
    \fill[white] svg{M86.3,186.2H70.9V79.1h15.4v48.4V186.2z}
                 svg{M108.9,79.1h41.6c39.6,0,57,28.3,57,53.6c0,27.5-21.5,53.6-56.8,53.6h-41.8V79.1z M124.3,172.4h24.5c34.9,0,42.9-26.5,42.9-39.7c0-21.5-13.7-39.7-43.7-39.7h-23.7V172.4z}
                 svg{M88.7,56.8c0,5.5-4.5,10.1-10.1,10.1c-5.6,0-10.1-4.6-10.1-10.1c0-5.6,4.5-10.1,10.1-10.1C84.2,46.7,88.7,51.3,88.7,56.8z};}}
\newcommand\orcid[1]{\href{https://orcid.org/#1}{\mbox{\scalerel*{
\begin{tikzpicture}[yscale=-1,transform shape]
\pic{orcidlogo};
\end{tikzpicture}
}{|}}}}
\begin{document}


\title{Odd-Parity Perturbations of Geometrically Regular Black Holes
with Hedgehog Scalar Hair}

\author{G.~G.~L. Nashed}\email{nashed@bue.edu.eg}

\affiliation{Centre for Theoretical Physics, The British University in Egypt,
P.O. Box 43, El Sherouk City, Cairo 11837, Egypt\\Centre for Space Research, North-West University,
Potchefstroom, South Africa}

\date{\today}
\begin{abstract}
In this paper, we study linear perturbations of a geometrically regular
black hole supported by a constrained $\mathrm{SO}(3)$ scalar triplet in
the hedgehog configuration. Since the individual background scalars depend
explicitly on the angular coordinates, the decoupling of polar and axial
perturbations cannot be concluded from the spherical form of the metric or
the energy--momentum tensor alone. To settle this point, we introduce the
polar and axial metric and triplet amplitudes simultaneously and derive the
corresponding projections of the radiative linearized Einstein--matter
equations for $\ell\geq2$. The polar equations contain only polar amplitudes,
whereas the axial equations contain only axial amplitudes; hence, the two
off-diagonal blocks vanish at linear order. The resulting odd-parity sector forms a closed Einstein--scalar system in
which a tangent perturbation of the scalar triplet is coupled to the
Regge--Wheeler metric amplitudes. We show that its scalar equation has a
second-order hyperbolic principal part whenever $K_Y\neq0$, and that, for
the nonlinear kinetic function supporting the background, $K_Y>0$
throughout the black-hole exterior. The coefficient of the direct
metric--scalar mixing is largest at the event horizon and decreases as
$r^{-4}$ at large distance, while the metric block in the vacuum limit
reproduces the standard Regge--Wheeler equation and potential; the scalar
kinetic coefficient simultaneously vanishes, so the full operator changes
rank. These results establish linear parity decoupling and the consistency of the
closed axial equations.  The odd sector can furthermore be reduced to two
coupled master variables, and the corresponding reduced quadratic action has
positive kinetic and radial-gradient eigenvalues at every finite exterior
radius for the regular branch with $K_Y>0$.  The scalar normalization tends to
zero asymptotically, however, and positivity of the full coupled potential
operator is not established; hence we do not claim a complete mode-stability
theorem.
\end{abstract}

\maketitle

\section{Introduction}
\label{sec:introduction}

The occurrence of singularities in gravitational collapse remains one of
the main indications that classical General Relativity (GR) cannot provide
a complete description of gravity at arbitrarily large curvatures. Under
broad physical assumptions, the singularity theorems imply geodesic
incompleteness \cite{Penrose:1964wq,Hawking:1973uf}. They do not, however,
identify the physical mechanism that should replace the singular region.
This has motivated the construction of black-hole geometries in which the
central divergence is replaced by a regular core while the essential
properties of a black hole are preserved.

The first well-known regular black-hole model was introduced by Bardeen
\cite{Bardeen:1968}. Since then, nonsingular geometries have been obtained
using nonlinear electrodynamics, effective matter distributions, modified
gravitational theories, and related extensions of GR
\cite{AyonBeato:1998ub,AyonBeato:1999rg,Bronnikov:2000vy,
Hayward:2005gi,Fan:2016hvf}. In many of these models, the Schwarzschild
singularity is replaced by a finite-curvature region that approaches a de
Sitter core. The regularity of the metric and curvature invariants is an
important requirement, but it is not sufficient by itself. The matter
configuration supporting the geometry must be treated consistently, and
the behavior of the solution under small perturbations must also be
examined.

Linear perturbation theory supplies the natural first step in this
direction. For a static and spherically symmetric background, metric
perturbations are conventionally expanded in spherical harmonics and
classified into polar and axial sectors
\cite{Regge:1957td,Zerilli:1970se,Moncrief:1974am}. In vacuum GR, the
odd-parity perturbations of the Schwarzschild geometry are described by
the Regge--Wheeler equation \cite{Regge:1957td}. Its effective potential
provides the standard basis for studying axial stability. The situation
becomes less direct when matter supports the background: gravitational
and matter perturbations can be coupled, additional modes may occur, and
the constraint structure can differ from that of the vacuum system.

In the present study, we consider the geometrically regular black hole
obtained from a constrained $\mathrm{SO}(3)$ scalar triplet and an
auxiliary three-form sector \cite{Bahamonde:2026bvh}. The triplet is
arranged in a hedgehog configuration, in which its internal orientation
follows the angular direction in physical space, while a Lagrange
multiplier fixes its modulus. The resulting energy--momentum tensor
supports an asymptotically flat black hole with a regular de Sitter core.
Unlike a radial scalar background, however, each member of the triplet
depends explicitly on $(\theta,\phi)$ even though the metric and the total
energy--momentum tensor are spherically symmetric.

This angular dependence is central to the perturbation problem. For an
ordinary radial scalar, scalar perturbations do not generally enter the
axial projections. In the hedgehog system, a perturbation tangent to the
internal two-sphere possesses an axial component and couples directly to
the Regge--Wheeler metric amplitudes. It is therefore not sufficient to
assume that odd and even perturbations decouple on the basis of the
spherical metric. The polar and axial variables must be introduced at the
same time, and the off-diagonal projections of the radiative linearized
operator must be calculated explicitly.

The constrained fields must also be handled with care. The perturbations
of the Lagrange multiplier, the auxiliary scalar, and the three-form are
retained until their field equations are imposed. The fixed-modulus
condition removes the normal triplet fluctuation, whereas the auxiliary
sector introduces no independent local radiative amplitude. This procedure
prevents the constraint structure from being inferred merely from the
absence of a second-order time derivative in one of the component
equations.

The main aim of this paper is to determine whether the polar and axial
perturbations of the hedgehog background decouple at linear order and to
obtain the resulting closed odd-parity equations. To achieve this aim, we
introduce the polar and axial metric and triplet amplitudes simultaneously.
We derive the polar Einstein--triplet projections, the axial projections,
and the two tangent-triplet equations. The calculation shows that each
polar projection depends only on polar amplitudes and each axial projection
depends only on axial amplitudes. Thus, both off-diagonal blocks of the
radiative Einstein--triplet operator vanish for $\ell\geq2$, while the
three-form constraint sector introduces no independent local radiative
amplitude.

We further examine the principal part of the tangent-scalar equation and
the radial behavior of the direct metric--scalar mixing coefficient. For
the nonlinear kinetic function supporting the regular background,
$K_Y>0$ gives the conventional sign of the bare scalar kinetic term and a
hyperbolic radial principal part in the exterior. The mixing coefficient
attains its largest exterior value at the event horizon and decreases as
$r^{-4}$ at large distance. This coefficient is not a canonically
normalized interaction strength, and these properties alone do not prove
the absence of ghosts, gradient instabilities, or unstable modes. As an
additional consistency test, the metric block in the matter-free limit
reproduces the standard Regge--Wheeler equation and potential. The scalar
kinetic coefficient vanishes in the same limit, so this is not a smooth
limit of the full coupled operator.

The structure of this paper is as follows. In
Sec.~\ref{sec:background}, we review the regular black-hole background and
develop the simultaneous polar--axial decomposition. In
Sec.~\ref{sec:complete_polar_system}, we derive the complete polar
projections. In Sec.~\ref{sec:odd_perturbations}, the closed odd-parity
Einstein--scalar equations are obtained. In Sec.~\ref{sec:odd_stability},
we discuss their gauge-invariant content, scalar principal part, radial
mixing coefficient, and Schwarzschild limit. Finally, in
Sec.~\ref{sec:conclusions}, we summarize the main results and state the
requirements for a complete mode-stability analysis. Supporting details
of the axial projections are given in the appendix.

\section{Background Geometry and Field Equations}
\label{sec:background}

We use the regular black-hole solution of Ref.~\cite{Bahamonde:2026bvh} as our background, with signature $(-,+,+,+)$ and $c=1$.

\subsection{Review of the Exact Solution}

The theory consists of Einstein gravity coupled to a constrained
$\mathrm{SO}(3)$ scalar triplet and an auxiliary three-form sector. Its action is
\begin{equation}
\begin{split}
S
=
\int d^4x\,\sqrt{-g}\,
\Bigg[
&
\frac{R}{16\pi G}
-
C\,K_b(Y)
+
\lambda
\left(
\delta_{IJ}\Phi^I\Phi^J-\eta^2
\right)
-
\frac{1}{3!}
\frac{\epsilon^{\mu\nu\rho\sigma}}{\sqrt{-g}}
A_{\mu\nu\rho}
\nabla_\sigma C
\Bigg],
\end{split}
\label{action}
\end{equation}
where
\begin{equation}
Y
=
\frac12
\delta_{IJ}
\nabla_\mu\Phi^I
\nabla^\mu\Phi^J
\label{Y}
\end{equation}
is the kinetic invariant of the scalar triplet.

From the three-form equation,
\begin{equation}
\nabla_\mu C=0,
\end{equation}
$C$ is constant. We denote this constant by $\rho_0$ and set
\begin{equation}
C=\rho_0,
\qquad
K(Y)=\rho_0 K_b(Y).
\label{Keff}
\end{equation}

Variation of the Lagrange multiplier gives
\begin{equation}
\Phi^I\Phi_I=\eta^2.
\label{constraint}
\end{equation}
We look for a static spherical solution by taking
\begin{equation}
\Phi^I=H(r)n^I(\theta,\phi),
\label{hedgehog}
\end{equation}
where
\begin{equation}
n^I=
(\sin\theta\cos\phi,
 \sin\theta\sin\phi,
 \cos\theta),
\qquad
n^In_I=1.
\end{equation}
For the metric, we use
\begin{equation}
ds^2
=
-A(r)dt^2
+\frac{dr^2}{A(r)}
+r^2
\left(
d\theta^2+\sin^2\theta\,d\phi^2
\right).
\label{metric}
\end{equation}

The constraint in Eq.~\eqref{constraint} fixes
\begin{equation}
H(r)=\eta,\quad \mbox{so that} \quad
Y=\frac{\eta^2}{r^2}.
\label{Yexact}
\end{equation}
The background scalar-field equation is
\begin{equation}
\nabla_\mu
\left(
K_Y\nabla^\mu\Phi^I
\right)
+
2\lambda\Phi^I
=
0.
\label{background_scalar_equation}
\end{equation}
For the hedgehog configuration $\Phi^I=\eta n^I$, the internal unit
vector satisfies
\begin{equation}
D^2 n^I=-2n^I.
\end{equation}
Projecting Eq.~\eqref{background_scalar_equation} along $n^I$ then gives
\begin{equation}
\bar\lambda
=
\frac{K_Y}{r^2},
\qquad
Y=\frac{\eta^2}{r^2}.
\label{background_lambda}
\end{equation}
Here $K_Y=dK/dY$ is evaluated on the background.
The Einstein equations then reduce to
\begin{equation}
m'(r)
=
4\pi r^2
K
\!\left(
\frac{\eta^2}{r^2}
\right),
\label{mass}
\end{equation}
with
\begin{equation}
A(r)
=
1-\frac{2Gm(r)}{r}.
\label{A}
\end{equation}

For the family
\begin{equation}
K_b(Y)
=
\left(
\frac{Y}{Y+\mu_*^2}
\right)^n,
\qquad
n>\frac32,
\end{equation}
the field equations admit asymptotically flat and geometrically regular
black-hole solutions \cite{Bahamonde:2026bvh}. We focus on the $n=3$
branch, for which the metric function is
\begin{equation}
A(r)
=
1-
\frac{4GM}{\pi r}
\left[
\arctan\!\left(\frac{r}{L}\right)
+
\frac{r}{L}
\frac{r^2/L^2-1}
{\left(1+r^2/L^2\right)^2}
\right],
\label{metric_function}
\end{equation}
where
\begin{equation}
L=\frac{\eta}{\mu_*}
\end{equation}
sets the size of the regular core. Regularity at the origin fixes the
relation
\begin{equation}
M
=
\frac{\pi^{2}}{4}\rho_{0}L^{3},
\label{MassDensityRelation}
\end{equation}
so that $M$, $\rho_0$, and $L$ are not independent.

The geometry has a de Sitter core,
\begin{equation}
A(r)
=
1-
\frac{8\pi G\rho_0}{3}r^2
+
\mathcal{O}(r^4),
\end{equation}
and approaches Schwarzschild at large radius,
\begin{equation}
A(r)
=
1-
\frac{2GM}{r}
+
\frac{32GML^3}{3\pi r^4}
+
\mathcal{O}(r^{-6}).
\label{asymptotic}
\end{equation}
Thus, the spacetime is asymptotically flat and free of curvature
singularities. The regularity discussed here is geometric: the metric and curvature
invariants remain finite at the center. The hedgehog scalar
configuration is nevertheless direction-dependent at $r=0$ and is not
a completely smooth matter configuration there. This issue does not
affect the exterior perturbation analysis performed in the present
work. The leading deviation from Schwarzschild appears only at
order $r^{-4}$, a property that will be useful when discussing the
asymptotic behavior of the perturbation potentials.

\subsection{Properties of the Background}
\label{subsec:background_properties}

For the perturbation analysis it is convenient to introduce the
dimensionless variables
\begin{equation}
x\equiv\frac{r}{L},
\qquad
\chi\equiv\frac{GM}{L},
\label{dimensionless_variables}
\end{equation}
where $\chi$ measures the ratio between the gravitational radius and the
size of the regular core. In terms of these variables, the metric
function becomes
\begin{equation}
A(x)
=
1-
\frac{4\chi}{\pi x}
\left[
\arctan x
+
\frac{x(x^2-1)}
{(1+x^2)^2}
\right].
\label{dimensionless_metric}
\end{equation}

\subsubsection{Horizon structure}

The Killing horizons are determined by the positive roots of
\begin{equation}
A(r_h)=0,
\label{horizon_condition}
\end{equation}
or equivalently,
\begin{equation}
A(x_h)=0.
\label{dimensionless_horizon_condition}
\end{equation}
Expressing the mass parameter as a function of the horizon radius gives
\begin{equation}
\chi(x_h)
=
\frac{\pi x_h}{4}
\left[
\arctan x_h
+
\frac{x_h(x_h^2-1)}
{(1+x_h^2)^2}
\right]^{-1},
\label{chi_horizon}
\end{equation}
which determines the horizon structure of the spacetime.

Depending on the value of $\chi$, three distinct configurations are
obtained~\cite{Bahamonde:2026bvh}. For
$\chi<\chi_{\mathrm{ext}}$, no event horizon exists and the geometry is
regular everywhere. At the critical value
\begin{equation}
\chi_{\mathrm{ext}}
\simeq
0.97686,
\qquad
x_{\mathrm{ext}}
\simeq
1.18924,
\label{extremal_values}
\end{equation}
the two horizons coincide, giving an extremal black hole. When
$\chi>\chi_{\mathrm{ext}}$, the solution has an inner Cauchy horizon
$r_-$ and an outer event horizon $r_+$, ordered as
\begin{equation}
r_-<r_+.
\label{two_horizons}
\end{equation}

Here we consider only the black-hole exterior,
\begin{equation}
r\geq r_+,
\end{equation}
and study the odd-parity perturbations in this region.


Although the individual components $\bar\Phi^I=\eta n^I(\theta,\phi)$
depend on the angles, the complete background is invariant under the
diagonal action of spatial rotations and internal $\mathrm{SO}(3)$
rotations.  Spherical symmetry of the metric and stress tensor alone
would not, however, be sufficient to establish parity decoupling. We
therefore retain both polar and axial amplitudes and derive the complete
linearized equations before restricting the subsequent discussion to
the axial block.

There is also a discrete symmetry that is important for the parity
classification.  Although we refer to the connected internal group as
$\mathrm{SO}(3)$, the action in Eq.~\eqref{action} contains the triplet
only through contractions with $\delta_{IJ}$.  It is consequently
invariant under the full internal $\mathrm{O}(3)$, including
$\Phi^I\mapsto-\Phi^I$.  Under the antipodal map ${\cal P}$ on the
two-sphere,
\begin{equation}
n^I({\cal P}\Omega)=-n^I(\Omega).
\end{equation}
The combined transformation
\begin{equation}
\widetilde{\cal P}
=({\cal P}\ \mathrm{on}\ S^2)
\circ(-\mathbf 1\ \mathrm{in\ internal\ space})
\label{combined_hedgehog_parity}
\end{equation}
therefore leaves the complete hedgehog background fixed.  The
linearized operator commutes with $\widetilde{\cal P}$.  Polar and axial
harmonics belong to its opposite eigenspaces, so a mixed block is
forbidden.  Because this symmetry statement depends on the particular
triplet action, rather than on the metric alone, we verify it directly
below by retaining all amplitudes in the field equations.

We use a $2+2$ harmonic decomposition. With $D_A$ and $\epsilon_{AB}$ denoting the
covariant derivative and Levi--Civita tensor of the unit two-sphere,
the axial harmonics are
\begin{equation}
X_A^{\ell m}=-\epsilon_A{}^B D_BY^{\ell m},\qquad
X_{AB}^{\ell m}=D_{(A}X_{B)}^{\ell m}.
\label{axial_harmonics_compact}
\end{equation}
They exist for $\ell\geq1$ and $\ell\geq2$, respectively. The
odd-parity metric perturbation is described by
$h_{aB}=h_aX_B^{\ell m}$ and $h_{AB}=h_2X_{AB}^{\ell m}$.

The most general scalar-triplet perturbation for a fixed $(\ell,m)$ can
be written as
\begin{equation}
\begin{split}
\delta\Phi^I=\eta\Big[&u(t,r)Y^{\ell m}n^I
+v(t,r)D_A Y^{\ell m}e^{AI}
\\
&+\psi(t,r)X_A^{\ell m}e^{AI}\Big],
\end{split}
\label{complete_triplet_perturbation}
\end{equation}
where $e_A{}^I=D_An^I$.  The linearized modulus constraint gives
$u=0$.  The amplitude $v$ is polar and $\psi$ is axial.  The metric is
decomposed simultaneously into the usual polar amplitudes
$H_{ab}Y^{\ell m}$, $j_aD_AY^{\ell m}$,
$r^2(K_g\Omega_{AB}Y^{\ell m}+G D_AD_BY^{\ell m})$, and the axial amplitudes
$h_aX_A^{\ell m}$ and $h_2X_{AB}^{\ell m}$.

We now perform the check required by the angular dependence of the
hedgehog, before choosing a polar or axial gauge.  Write the tangent
triplet perturbation and the general metric perturbation as
\begin{align}
\zeta_A&=vD_AY^{\ell m}+\psi X_A^{\ell m},
\quad \delta\Phi^I=\eta\zeta_Ae^{AI},
\nonumber\\
p_{ab}&=H_{ab}Y^{\ell m},\quad
p_{aA}=j_aD_AY^{\ell m}+h_aX_A^{\ell m},
\nonumber\\
p_{AB}&=r^2K_g\Omega_{AB}Y^{\ell m}
+r^2G D_AD_BY^{\ell m}+h_2X_{AB}^{\ell m}.
\label{unrestricted_metric_triplet_ansatz}
\end{align}
No Regge--Wheeler gauge has been imposed in
Eq.~\eqref{unrestricted_metric_triplet_ansatz}.  Direct variation of
$Y$ gives
\begin{equation}
\delta Y=
\frac{\eta^2}{r^2}D_A\zeta^A
-\frac{\eta^2}{2r^4}\Omega^{AB}p_{AB}.
\label{main_general_delta_Y}
\end{equation}
Because $D_AX^A=0$ and $\Omega^{AB}X_{AB}=0$, its axial part vanishes
identically:
\begin{equation}
\delta Y[q_{\mathrm{ax}}]=0.
\label{main_axial_delta_Y_zero}
\end{equation}
The polar part is proportional to $Y^{\ell m}$.  Hence $\delta K$ and
$\delta K_Y$ are polar scalars and cannot supply an axial source.

Varying the complete energy--momentum tensor, without setting either
$v$ or $\psi$ to zero, gives
\begin{align}
\delta T_{ab}&=-\delta K\,\bar g_{ab}-Kp_{ab},
\nonumber\\
\delta T_{aA}&=\eta^2K_Y\partial_a\zeta_A-Kp_{aA},
\nonumber\\
\delta T_{AB}&=\eta^2K_Y(D_A\zeta_B+D_B\zeta_A)-Kp_{AB}
\nonumber\\
&\quad+(\eta^2\delta K_Y-r^2\delta K)\Omega_{AB}.
\label{main_complete_delta_T}
\end{align}
Equation~\eqref{main_complete_delta_T} separates directly into
\begin{align}
\delta T_{ab}^{\mathrm{ax}}&=0,
\qquad
\delta T_{aA}^{\mathrm{ax}}
=(\eta^2K_Y\partial_a\psi-Kh_a)X_A^{\ell m},
\nonumber\\
\delta T_{AB}^{\mathrm{ax}}
&=(2\eta^2K_Y\psi-Kh_2)X_{AB}^{\ell m},
\label{main_axial_delta_T}
\end{align}
and
\begin{align}
\delta T_{ab}^{\mathrm{pol}}
&=-\delta K\,\bar g_{ab}-KH_{ab}Y^{\ell m},
\nonumber\\
\delta T_{aA}^{\mathrm{pol}}
&=(\eta^2K_Y\partial_av-Kj_a)D_AY^{\ell m},
\nonumber\\
\delta T_{AB}^{\mathrm{pol}}
&=2\eta^2K_YvD_AD_BY^{\ell m}
-Kr^2K_g\Omega_{AB}Y^{\ell m}
\nonumber\\
&\quad-Kr^2GD_AD_BY^{\ell m}
+(\eta^2\delta K_Y-r^2\delta K)\Omega_{AB}.
\label{main_polar_delta_T}
\end{align}
Thus the matter calculation itself, rather than an assumption based on
$\bar T_{\mu\nu}$, contains no mixed polar--axial projection.

The same test is required for the triplet equation.  Projecting its
first variation along $e_A{}^I$ while retaining both $v$ and $\psi$
gives the one-form
\begin{equation}
\delta{\cal S}_A
={\cal S}_{\mathrm{pol}}D_AY^{\ell m}
+{\cal S}_{\mathrm{ax}}X_A^{\ell m}.
\label{main_scalar_parity_split}
\end{equation}
The variation $\delta K_Y$ contributes only
$\eta D_A\delta K_Y/r^2$, while
$e_{AI}\delta\lambda\,\bar\Phi^I=0$.  Moreover,
$D^2D_AY^{\ell m}=-(L_\ell-1)D_AY^{\ell m}$ and
$D^2X_A^{\ell m}=-(L_\ell-1)X_A^{\ell m}$.  Therefore the angular
operator preserves the two terms in Eq.~\eqref{main_scalar_parity_split}
separately.  In particular, its axial coefficient is
most compactly expressed using the gauge-invariant amplitude
\begin{equation}
\widehat\psi\equiv\psi-\frac{h_2}{2r^2}.
\label{main_gauge_invariant_psi}
\end{equation}
Indeed, under an axial diffeomorphism with
$\xi_A=\Lambda X_A^{\ell m}$ one has
$\psi\to\psi-\Lambda/r^2$ and $h_2\to h_2-2\Lambda$.
Before any gauge choice, the axial coefficient is therefore
\begin{align}
{\cal S}_{\mathrm{ax}}={}&
-\partial_t\!\left[\frac{r^2K_Y}{A}
\left(\dot\psi-\frac{h_0}{r^2}\right)\right]
+\partial_r\!\left[r^2AK_Y
\left(\psi'-\frac{h_1}{r^2}\right)\right]
\nonumber\\
&-\lambda_\ell K_Y\widehat\psi,
\label{main_simultaneous_axial_scalar}
\end{align}
which contains no polar amplitude.  Conversely,
${\cal S}_{\mathrm{pol}}$ contains $v$ and the polar metric and
auxiliary amplitudes, but no $\psi$, $h_a$, or $h_2$.

Finally, the linearized Einstein operator on the spherical metric
preserves the polar and axial tensor-harmonic subspaces before gauge
fixing. Combining this geometric fact with
Eqs.~\eqref{main_complete_delta_T}--\eqref{main_scalar_parity_split}
gives the four direct cross-block tests
\begin{align}
{\cal P}_{\mathrm{pol}}\delta E[q_{\mathrm{ax}}]&=0,
&{\cal P}_{\mathrm{ax}}\delta E[q_{\mathrm{pol}}]&=0,
\nonumber\\
{\cal P}_{\mathrm{pol}}\delta{\cal S}[q_{\mathrm{ax}}]&=0,
&{\cal P}_{\mathrm{ax}}\delta{\cal S}[q_{\mathrm{pol}}]&=0,
\label{main_four_cross_block_tests}
\end{align}
where $\delta E_{\mu\nu}=\delta G_{\mu\nu}-8\pi G\delta T_{\mu\nu}$.
These identities hold for every $(\ell,m)$ and are obtained before
imposing either polar or axial Regge--Wheeler gauge.  The constraint
sets $u=0$, the normal triplet equation fixes $\delta\lambda$, and the
three-form equation gives $\delta C=0$ for radiative multipoles; none
of them creates a cross-block term.

The calculation consequently gives
\begin{equation}
\delta Y_{\mathrm{ax}}=0,
\qquad
\delta T_{\mu\nu}[q_{\mathrm{ax}}]
=\left\{0,\mathcal T_aX_A,\mathcal T X_{AB}\right\},
\label{explicit_parity_check_summary}
\end{equation}
where the three entries refer respectively to the $(ab)$, $(aA)$ and
$(AB)$ components.  Conversely, the perturbations generated by
$q_{\mathrm{pol}}$ contain only $Y^{\ell m}$, $D_AY^{\ell m}$ and
polar tensor harmonics. The explicit polar Einstein and tangent-triplet
projections are displayed next in
Sec.~\ref{sec:complete_polar_system}. None of those equations contains
$h_0$, $h_1$, $h_2$ or $\psi$, while the axial projections derived below
contain none of $H_0$, $H_1$, $H_2$, $K_g$ or $v$.  Consequently, for
the present action and background, the radiative Einstein--triplet
operator for $\ell\geq2$ takes
the block-diagonal form
\begin{equation}
\begin{pmatrix}
\mathcal L_{\mathrm{pol}}&0\\
0&\mathcal L_{\mathrm{ax}}
\end{pmatrix}
\begin{pmatrix}q_{\mathrm{pol}}\\q_{\mathrm{ax}}\end{pmatrix}=0.
\label{verified_linear_parity_blocks}
\end{equation}
This conclusion follows from the displayed polar and axial equations,
not from spherical symmetry of the metric alone. The three-form
constraint sector is not an additional radiative block and introduces
no independent local radiative amplitude.

Having established the vanishing cross-blocks, the axial tangent
perturbation can be written as
\begin{equation}
\delta\Phi^I=\eta\psi(t,r)X_A^{\ell m}e^{AI}.
\label{compact_triplet_perturbation}
\end{equation}
Under an axial diffeomorphism
$\xi_A=\Lambda(t,r)X_A^{\ell m}$, its amplitude transforms as
\begin{equation}
\psi\longrightarrow\psi-\frac{\Lambda}{r^2}.
\label{psi_gauge_transformation}
\end{equation}
The perturbations $\delta\lambda$ and $\delta C$ are ordinary
scalars on $S^2$ and therefore have no axial harmonic. Moreover, the
linearized three-form equation gives $\nabla_\mu\delta C=0$; hence
the three-form sector supplies no independent local axial
amplitude. The radiative analysis below is restricted to $\ell\geq2$.

\section{Polar-Parity Perturbations}
\label{sec:complete_polar_system}
\begin{widetext}

For the radiative multipoles $\ell\geq2$, impose the polar
Regge--Wheeler gauge and write
\begin{align}
p_{tt}&=AH_0Y^{\ell m},
&p_{tr}&=H_1Y^{\ell m},
&p_{rr}&=\frac{H_2}{A}Y^{\ell m},
\nonumber\\
p_{AB}&=r^2K_g\Omega_{AB}Y^{\ell m},
&\delta\Phi^I_{\mathrm{pol}}
&=\eta vD_AY^{\ell m}e^{AI}.
\label{polar_rw_ansatz_appendix}
\end{align}
All amplitudes in this section depend on $(t,r)$.  The three-form
equation gives $\nabla_\mu\delta C=0$; hence $\delta C=0$ for
$\ell\geq2$.  It is convenient to define
\begin{align}
\delta Y&=\mathcal Y Y^{\ell m},
&\mathcal Y&=-\frac{\eta^2}{r^2}(L_\ell v+K_g),
\nonumber\\
\delta K&=\mathcal K Y^{\ell m},
&\mathcal K&=K_Y\mathcal Y,
\nonumber\\
\delta K_Y&=\mathcal K_Y Y^{\ell m},
&\mathcal K_Y&=K_{YY}\mathcal Y.
\label{polar_matter_variations_appendix}
\end{align}

The polar matter projections obtained before using any Einstein
equation are
\begin{align}
\delta T_{tt}^{\mathrm{pol}}
&=A(\mathcal K-KH_0)Y^{\ell m},
&
\delta T_{tr}^{\mathrm{pol}}
&=-KH_1Y^{\ell m},
\nonumber\\
\delta T_{rr}^{\mathrm{pol}}
&=-\frac{\mathcal K+KH_2}{A}Y^{\ell m},
&
\delta T_{tA}^{\mathrm{pol}}
&=\eta^2K_Y\dot vD_AY^{\ell m},
\nonumber\\
\delta T_{rA}^{\mathrm{pol}}
&=\eta^2K_Yv'D_AY^{\ell m},
\label{polar_matter_components_appendix}
\end{align}
and
\begin{align}
\delta T_{AB}^{\mathrm{pol}}
={}&\mathcal T_\Omega\Omega_{AB}Y^{\ell m}
+2\eta^2K_YvY_{AB}^{\ell m},
\nonumber\\
\mathcal T_\Omega
={}&-\eta^2K_YL_\ell v-r^2KK_g
+\eta^2\mathcal K_Y-r^2\mathcal K,
\label{polar_matter_angular_appendix}
\end{align}
where
\begin{equation}
Y_{AB}^{\ell m}
=D_AD_BY^{\ell m}+\frac{L_\ell}{2}\Omega_{AB}Y^{\ell m} \quad \mbox{is trace free.}
\end{equation}

Direct variation of the Einstein tensor gives the following complete
polar projections.  The $(tt)$ equation is
\begin{align}
\frac{A}{2r^2}\Big[&L_\ell(H_2+K_g)
-r^2(2AK_g''+A'K_g')
\nonumber\\
&+2r(AH_2'-3AK_g'+A'H_0+A'H_2)
\nonumber\\
&+2AH_0+2AH_2-2H_0-2K_g\Big]
=8\pi GA(\mathcal K-KH_0).
\label{polar_Einstein_tt_appendix}
\end{align}
The $(tr)$ equation is
\begin{align}
-\dot K_g'
+\frac{A'}{2A}\dot K_g
+\frac{A'H_1+\dot H_2-\dot K_g}{r}
+\frac{L_\ell+2A-2}{2r^2}H_1
=-8\pi GKH_1.
\label{polar_Einstein_tr_appendix}
\end{align}
The $(rr)$ equation is
\begin{align}
&\frac{1}{2r^2A^2}\Big[r^2AA'K_g'-2r^2\ddot K_g
+2rA^2(-H_0'+K_g')
\nonumber\\
&+4rA\dot H_1
+A\{L_\ell H_0-L_\ell K_g-2H_2+2K_g\}\Big]
=-\frac{8\pi G}{A}(\mathcal K+KH_2).
\label{polar_Einstein_rr_appendix}
\end{align}
The polar vector equations are
\begin{align}
\frac12(AH_1'+A'H_1-\dot H_2-\dot K_g)
&=8\pi G\eta^2K_Y\dot v,
\label{polar_Einstein_tA_appendix}\\
\frac{1}{4rA}\Big[2rA(H_0'-K_g')
+r\{A'(H_0+H_2)-2\dot H_1\}
\nonumber\\
+2A(-H_0+H_2)\Big]
&=8\pi G\eta^2K_Yv'.
\label{polar_Einstein_rA_appendix}
\end{align}
The trace-free angular equation is particularly simple:
\begin{equation}
\frac12(H_0-H_2)=16\pi G\eta^2K_Yv.
\label{polar_Einstein_TF_appendix}
\end{equation}
Finally, the angular trace equation is
\begin{equation}
\mathcal G_\Omega=8\pi G\mathcal T_\Omega,
\label{polar_Einstein_trace_appendix}
\end{equation}
where
\begin{align}
\mathcal G_\Omega
=\frac{1}{4A}\Big[&2r^2A'\dot H_1-2r^2\ddot H_2
-2r^2\ddot K_g
\nonumber\\
&+A\{L_\ell(H_0-H_2)-2r^2AH_0''
+2r^2AK_g''
\nonumber\\
&-2r^2H_2A''+2r^2K_gA''
-3r^2A'H_0'-r^2A'H_2'
\nonumber\\
&+2r^2A'K_g'+4r^2\dot H_1'
-2rAH_0'-2rAH_2'
\nonumber\\
&+4rAK_g'-4rH_2A'+4rK_gA'
+4r\dot H_1\}\Big].
\label{polar_GOmega_appendix}
\end{align}

The polar tangent projection of the triplet equation is
\begin{align}
&-\frac{K_Y}{A}\ddot v
+AK_Yv''
+\left(AK_Y'+A'K_Y+\frac{2AK_Y}{r}\right)v'
\nonumber\\
&-\frac{\lambda_\ell K_Y}{r^2}v
+\frac{K_Y}{2r^2}(H_2-H_0)
+\frac{\mathcal K_Y}{r^2}=0.
\label{polar_scalar_equation_appendix}
\end{align}
The normal triplet projection fixes the multiplier perturbation:
\begin{equation}
\delta\lambda
=\frac{1}{r^2}
\left[\mathcal K_Y-K_Y(L_\ell v+K_g)\right]Y^{\ell m}.
\label{polar_delta_lambda_appendix}
\end{equation}
It introduces no additional local amplitude into
Eqs.~\eqref{polar_Einstein_tt_appendix}--
\eqref{polar_scalar_equation_appendix}.
\end{widetext}

\section{Odd-Parity Perturbations}
\label{sec:odd_perturbations}

We now evaluate the axial block of the metric and matter equations.
The simultaneous matter calculation and the four cross-block tests
have been given above as part of the general perturbation development.
The simultaneous derivation establishes linear parity
decoupling \cite{Nashed:2021gkp}. The equations below support
limited checks of the scalar principal part and of the Schwarzschild
vacuum limit, but not a proof of mode stability.

\subsection{Axial Metric Perturbations}

To calculate the axial metric projections, we retain the axial metric
harmonics. Since
the odd tensor harmonics $X_{AB}^{\ell m}$ vanish for $\ell=0$ and
$\ell=1$, only multipoles with $\ell\ge2$ contribute to the radiative
axial perturbations considered here.

For each multipole $(\ell,m)$, the odd-parity metric perturbation is
decomposed as
\begin{align}
h^{\mathrm {odd}}_{ab}
&=0,
\quad
h^{\mathrm { odd}}_{aB}
=
h_a(t,r)\,
X^{\ell m}_B,
\label{odd2}
\\
h^{\mathrm { odd}}_{AB}
&=
h_2(t,r)\,
X^{\ell m}_{AB},
\label{odd3}
\end{align}
where
\begin{equation}
h_a=(h_0,h_1)
\end{equation}
is a covector on the orbit space $M^2$, while $h_2(t,r)$ denotes the
coefficient of the odd tensor harmonic. Since the components
$h_{ab}$ behave as scalars on the two-sphere, they belong only to the
even-parity sector and therefore vanish in the axial decomposition.

In component form, the metric perturbation takes the form
\begin{equation}
h^{\mathrm {odd}}_{\mu\nu}
=
\sum_{\ell,m}
\left(
\begin{array}{cccc}
0
&
0
&
h_0X_\theta
&
h_0X_\phi
\\
0
&
0
&
h_1X_\theta
&
h_1X_\phi
\\
h_0X_\theta
&
h_1X_\theta
&
h_2X_{\theta\theta}
&
h_2X_{\theta\phi}
\\
h_0X_\phi
&
h_1X_\phi
&
h_2X_{\theta\phi}
&
h_2X_{\phi\phi}
\end{array}
\right),
\label{oddmetric}
\end{equation}
where the multipole indices have been omitted for simplicity.

The odd-parity perturbations are still subject to infinitesimal
coordinate transformations,
\begin{equation}
x^\mu\rightarrow x^\mu+\xi^\mu,
\label{odd_coordinate_transformation}
\end{equation}
under which
\begin{equation}
h_{\mu\nu}
\rightarrow
h_{\mu\nu}
-\bar\nabla_\mu\xi_\nu
-\bar\nabla_\nu\xi_\mu,
\label{metric_gauge_transformation}
\end{equation}
with $\bar\nabla_\mu$ denoting the covariant derivative associated with
the background metric.

For odd parity, the gauge vector is chosen as
\begin{equation}
\xi_a=0,
\qquad
\xi_A=\Lambda(t,r)X_A^{\ell m},
\label{odd_gauge_vector_metric}
\end{equation}
where $\Lambda(t,r)$ is arbitrary. The metric amplitudes then transform
according to
\begin{align}
h_a
&\rightarrow
h_a-D_a\Lambda
+\frac{2}{r}(D_ar)\Lambda,
\label{odd_ha_transformation}
\\
h_2
&\rightarrow
h_2-2\Lambda.
\label{odd_h2_transformation}
\end{align}
Choosing
\begin{equation}
\Lambda=\frac{h_2}{2},
\label{rw_gauge_choice}
\end{equation}
eliminates the tensor amplitude,
\begin{equation}
h_2=0,
\label{regge_wheeler_gauge}
\end{equation}
which is the Regge--Wheeler gauge. The odd-parity metric perturbation
therefore becomes
\begin{equation}
h^{\mathrm {odd}}_{\mu\nu}
=
\sum_{\ell,m}
\left(
\begin{array}{cccc}
0
&
0
&
h_0X_\theta
&
h_0X_\phi
\\
0
&
0
&
h_1X_\theta
&
h_1X_\phi
\\
h_0X_\theta
&
h_1X_\theta
&
0
&
0
\\
h_0X_\phi
&
h_1X_\phi
&
0
&
0
\end{array}
\right),
\label{RWmetric}
\end{equation}
so that the axial metric sector is completely specified by the two
functions
\begin{equation}
h_0(t,r),
\qquad
h_1(t,r).
\end{equation}

\subsection{Odd-Parity Matter Perturbations}

The matter sector consists of the constrained scalar triplet, the Lagrange multiplier, and the auxiliary three-form introduced in Sec.~\ref{sec:background}. Here, we retain only the odd-parity perturbations that couple to the axial gravitational sector.
\subsubsection{Constrained Scalar Triplet}

The background scalar field is
\begin{equation}
\bar{\Phi}^I=\eta n^I,
\qquad
n_In^I=1,
\end{equation}
where $n^I$ denotes the unit vector in the internal space. As shown in
Sec.~\ref{sec:background}, the fixed-modulus constraint
\begin{equation}
\Phi_I\Phi^I=\eta^2
\end{equation}
removes the normal perturbation, leaving only fluctuations tangent to
the vacuum manifold,
\begin{equation}
\delta\Phi^I
=
\eta\,
\psi_Ae^{AI},\quad \mbox{with} \quad
e_A{}^{I}=D_An^I.
\end{equation}

Restricting to odd parity,
\begin{equation}
\psi_A
=
\sum_{\ell,m}
\psi_{\ell m}(t,r)
X_A^{\ell m},
\end{equation}
so that
\begin{equation}
\delta\Phi^I
=
\eta
\sum_{\ell,m}
\psi_{\ell m}(t,r)
X_A^{\ell m}
e^{AI}.
\label{oddscalar}
\end{equation}
Since $X_A^{\ell m}$ vanishes for $\ell=0$, the odd-parity scalar
perturbation starts at $\ell=1$. The equations derived below show that
$\psi_{\ell m}$ has a second-order principal part wherever
$K_Y\neq0$. Whether it represents an independent canonical degree of
freedom in the reduced coupled system can be determined only after an
action-level analysis of the constraints.
\paragraph{Auxiliary fields and gauge-invariant variables.}

The Lagrange multiplier and auxiliary fields are expanded according to
\begin{equation}
\lambda
=
\bar{\lambda}
+\delta\lambda,\qquad
A_{\mu\nu\rho}
=
\bar A_{\mu\nu\rho}
+
\delta A_{\mu\nu\rho},
\end{equation}
and
\begin{equation}
C
=
\rho_0+\delta C.
\end{equation}
Since $\delta\lambda$ is a scalar on the two-sphere, it belongs to the
even-parity sector and therefore does not contribute directly to the
axial perturbations. Likewise, the auxiliary three-form carries no
local propagating degrees of freedom in four dimensions
\cite{Aurilia:1980,Henneaux:1984}. Its perturbations,
together with $\delta C$, are retained until their constraint equations
are taken into account. They introduce no independent local axial
amplitude in the closed Einstein--scalar equations used below.

The odd-parity scalar perturbation transforms according to
Eq.~\eqref{psi_gauge_transformation}, while the metric amplitudes obey
Eqs.~\eqref{odd_ha_transformation} and
\eqref{odd_h2_transformation}. Consequently, the combination
\begin{equation}
\widehat{\psi}
=
\psi-\frac{h_2}{2r^2}
\label{gauge_invariant_psi}
\end{equation}
is gauge invariant.

A second gauge-invariant quantity is
\begin{equation}
Q_a
=
D_a\psi-\frac{h_a}{r^2},
\label{gauge_invariant_Qa}
\end{equation}
whose invariance follows directly from the transformation laws of
$\psi$ and $h_a$. In the Regge--Wheeler gauge,
\begin{equation}
h_2=0,\quad \mbox{one simply has} \quad
\widehat{\psi}=\psi.
\end{equation}
\subsection{Linearized Field Equations}

Substituting the odd-parity perturbations into the Einstein and matter
field equations and retaining terms linear in the perturbation
amplitudes yields the coupled Einstein--scalar system,
\begin{equation}
\delta G_{\mu\nu}
=
8\pi G\,\delta T_{\mu\nu},
\label{LinearizedEinsteinEquations}
\end{equation}
where only the $(tA)$, $(rA)$, and trace-free $(AB)$ components
contribute. Throughout this subsection we work in the
Regge--Wheeler gauge,
\begin{equation}
h_{tA}=h_0X_A^{\ell m},
\qquad
h_{rA}=h_1X_A^{\ell m},
\label{OddMetricRWComponents}
\end{equation}
and introduce
\begin{equation}
L_\ell=\ell(\ell+1),
\qquad
\lambda_\ell=(\ell-1)(\ell+2)=L_\ell-2.
\label{OddAngularEigenvalues}
\end{equation}

For the constrained scalar triplet,
\begin{equation}
T_{\mu\nu}
=
K_Y\nabla_\mu\Phi^I\nabla_\nu\Phi_I
-
Kg_{\mu\nu},
\label{MatterEnergyMomentum}
\end{equation}
with
\begin{equation}
Y=\frac{\eta^2}{r^2},
\end{equation}
the background Einstein equations imply
\begin{align}
\frac{rA'+A-1}{r^2}
&=
-8\pi GK,
\label{BackgroundEinsteinTTIdentity}
\\
\frac{A''}{2}
+
\frac{A'}{r}
&=
8\pi G
\left(
\frac{\eta^2K_Y}{r^2}
-
K
\right),
\label{BackgroundEinsteinAngularIdentity}
\end{align}
which are useful for simplifying the perturbation equations.

For the odd-parity scalar perturbation
\begin{equation}
\delta\Phi^I
=
\eta\psi
X_A^{\ell m}
e^{AI},
\label{OddScalarForFieldEquations}
\end{equation}
the first-order variation of the kinetic invariant vanishes,
\begin{equation}
\delta Y_{\mathrm{odd}}=0.
\label{OddDeltaYZeroFieldEquations}
\end{equation}
Consequently, the nonvanishing components of the perturbed
energy--momentum tensor are
\begin{align}
\delta T_{tA}
&=
\left(
\eta^2K_Y\dot\psi
-
Kh_0
\right)
X_A^{\ell m},
\label{ExplicitDeltaTtA}
\\
\delta T_{rA}
&=
\left(
\eta^2K_Y\psi'
-
Kh_1
\right)
X_A^{\ell m},
\label{ExplicitDeltaTrA}
\\
\delta T_{AB}^{\mathrm{TF}}
&=
2\eta^2K_Y\psi
X_{AB}^{\ell m}.
\label{ExplicitDeltaTAB}
\end{align}
where an overdot and a prime denote differentiation with respect to
$t$ and $r$.

The corresponding odd-parity components of the Einstein tensor are
\begin{align}
\delta G_{tA}
={}&
\frac12
\Bigg[
-Ah_0''
+
A\dot h_1'
+
\frac{2A}{r}\dot h_1
\nonumber\\
&
+
\left(
A''
+
\frac{2A'}{r}
+
\frac{L_\ell+2A-2}{r^2}
\right)
h_0
\Bigg]
X_A^{\ell m},
\label{ExplicitDeltaGtA}
\\
\delta G_{rA}
={}&
\frac12
\Bigg[
\frac{\ddot h_1}{A}
-
\frac{\dot h_0'}{A}
+
\frac{2\dot h_0}{rA}
\nonumber\\
&
+
\left(
A''
+
\frac{2A'}{r}
+
\frac{\lambda_\ell}{r^2}
\right)
h_1
\Bigg]
X_A^{\ell m},
\label{ExplicitDeltaGrA}
\\
\delta G_{AB}^{\mathrm{TF}}
={}&
\left(
Ah_1'
+
A'h_1
-
\frac{\dot h_0}{A}
\right)
X_{AB}^{\ell m}.
\label{ExplicitDeltaGAB}
\end{align}

Projecting the Einstein equations onto the odd vector and tensor
harmonics gives
\begin{align}
&
-Ah_0''
+
A\dot h_1'
+
\frac{2A}{r}\dot h_1
+
\left(
A''
+
\frac{2A'}{r}
+
\frac{L_\ell+2A-2}{r^2}
\right)
h_0
\nonumber\\
&
\hspace{2cm}
=
16\pi G
\left(
\eta^2K_Y\dot\psi
-
Kh_0
\right),
\label{ExplicitOddEinsteinTA}
\end{align}

\begin{align}
&
\frac{\ddot h_1}{A}
-
\frac{\dot h_0'}{A}
+
\frac{2\dot h_0}{rA}
+
\left(
A''
+
\frac{2A'}{r}
+
\frac{\lambda_\ell}{r^2}
\right)
h_1
\nonumber\\
&
\hspace{2cm}
=
16\pi G
\left(
\eta^2K_Y\psi'
-
Kh_1
\right),
\label{ExplicitOddEinsteinRA}
\end{align}

and
\begin{equation}
Ah_1'
+
A'h_1
-
\frac{\dot h_0}{A}
=
16\pi G\eta^2K_Y\psi.
\label{ExplicitOddEinsteinAB}
\end{equation}

The tangent projection of the scalar field equation becomes
\begin{align}
&-\partial_t
\left[
\frac{r^2K_Y}{A}
\left(
\dot\psi
-
\frac{h_0}{r^2}
\right)
\right]
+
\partial_r
\left[
r^2AK_Y
\left(
\psi'
-
\frac{h_1}{r^2}
\right)
\right]\nonumber\\
&-
\lambda_\ell K_Y\psi
=
0.
\label{ExplicitOddScalarEquation}
\end{align}

Introducing the gauge-invariant combination
\begin{equation}
Q_a
=
D_a\psi
-
\frac{h_a}{r^2},
\label{EquivalentQaExpression}
\end{equation}
the scalar equation can be written covariantly as
\begin{equation}
D_a
\left(
r^2K_YQ^a
\right)
-
\lambda_\ell K_Y\widehat\psi
=
0,
\label{CovariantOddScalarEquation}
\end{equation}
which, in the Regge--Wheeler gauge, reduces to
\begin{equation}
D_a
\left[
r^2K_Y
\left(
D^a\psi
-
\frac{h^a}{r^2}
\right)
\right]
-
\lambda_\ell K_Y\psi
=
0.
\label{CovariantOddScalarEquationRW}
\end{equation}

Equations~\eqref{ExplicitOddEinsteinTA}--\eqref{ExplicitOddEinsteinAB},
together with the tangent scalar equation derived above, are the axial
projections of the Einstein--scalar equations for each multipole
$\ell\geq2$. The complete polar equations in
Sec.~\ref{sec:complete_polar_system} contain no axial amplitude, so
these axial equations form a closed linear subsystem.
They are not four independent equations.  To display the corresponding
consistency identity, define
$\mathcal E_{\mu\nu}=G_{\mu\nu}-8\pi G T_{\mu\nu}$ and let
$\mathcal E_I=\nabla_\mu(K_Y\nabla^\mu\Phi_I)+2\lambda\Phi_I$
denote the triplet Euler--Lagrange expression after imposing the
auxiliary equations. Diffeomorphism invariance gives the exact Noether
identity
\begin{equation}
\nabla^\mu\mathcal E_{\mu\nu}
=-8\pi G\,\mathcal E_I\nabla_\nu\Phi^I.
\label{axial_noether_identity}
\end{equation}
On the background, its linearized axial projection is
\begin{equation}
D^a\delta\mathcal E_a
+\frac{2D^ar}{r}\delta\mathcal E_a
-\frac{\lambda_\ell}{2r^2}\delta\mathcal E_2
=-8\pi G\eta\,e_A{}^I\delta\mathcal E_I,
\label{projected_axial_noether_identity}
\end{equation}
where
$\delta\mathcal E_{aA}=\delta\mathcal E_aX_A^{\ell m}$ and
$\delta\mathcal E_{AB}=\delta\mathcal E_2X_{AB}^{\ell m}$; the common
$X_A^{\ell m}$ factor is understood on the right-hand side.  We used
$D^BX_{AB}^{\ell m}=-(\lambda_\ell/2)X_A^{\ell m}$. Thus, once the
three axial Einstein projections hold, the tangent-triplet equation
follows (and conversely one Einstein projection is propagated by the
others and the triplet equation). This verifies explicitly that the
closed subsystem is not overdetermined.
{In the original $(h_0,h_1,\psi)$ formulation, the $(tA)$ equation
contains radial derivatives of $h_0$, while
Eq.~\eqref{ExplicitOddEinsteinAB} provides a first-order reconstruction
relation for $\dot h_0$.  These statements refer to the unreduced projected
variables.  As shown below, after introducing the gauge-invariant curl
$\Pi$ the projected Einstein equations can instead be rearranged into
algebraic reconstruction formulae for both $h_0$ and $h_1$ in terms of the
master variables.  The tangent scalar equation contains a second-order time
derivative whenever $K_Y\neq0$, whereas the Lagrange-multiplier and
three-form sectors introduce no independent local axial amplitudes.  The
field-equation master reduction determines the two coupled dynamical
amplitudes; the corresponding quadratic-action reduction and canonical
normalization are carried out below.}

\section{Odd-Parity Dynamical Content and Stability Criteria}
\label{sec:odd_stability}

In this section, we work with the closed axial equations derived in
Sec.~\ref{sec:odd_perturbations}, while keeping separate the question
of parity decoupling from the stronger question of mode stability.

\subsection{Gauge-invariant variables and closed field equations}
\label{subsec:odd_gi_system}

For each radiative multipole $\ell\geq2$, define
\begin{equation}
 L_\ell=\ell(\ell+1),
 \qquad
 \lambda_\ell=(\ell-1)(\ell+2)=L_\ell-2.
 \label{odd_angular_eigenvalues_stability}
\end{equation}
Before gauge fixing, the axial metric perturbation is described by $h_a$ and
$h_2$.  Under the odd gauge transformation generated by $\Lambda(t,r)$, the
combination
\begin{equation}
 k_a=h_a-\frac{1}{2}D_a h_2+\frac{D_a r}{r}h_2
 \label{odd_gauge_invariant_ka_stability}
\end{equation}
is invariant.  The gauge-invariant tangent perturbation and its covariant
derivative are
\begin{equation}
 \widehat{\psi}=\psi-\frac{h_2}{2r^2},
 \qquad
 Q_a=D_a\psi-\frac{h_a}{r^2}.
 \label{odd_gauge_invariant_scalar_variables_stability}
\end{equation}
The second expression is gauge invariant under the conventions adopted in Sec.~\ref{sec:odd_perturbations}. In the Regge--Wheeler gauge, where $h_2=0$, one has $k_a=h_a$ and $\widehat{\psi}=\psi$.

The gauge-invariant gravitational curl is
\begin{equation}
 \Pi=\frac{1}{r^2}\epsilon^{ab}D_a\left(\frac{k_b}{r^2}\right),
 \label{odd_gauge_invariant_pi_stability}
\end{equation}
where the overall sign depends only on the orientation chosen for
$\epsilon^{ab}$.  Thus $h_1$ is a convenient Regge--Wheeler-gauge amplitude,
but it is not the invariant gravitational variable before gauge fixing.  The
quantity $\Pi$ gives a gauge-invariant description of the axial metric
perturbation. {The gauge-invariant curl $\Pi$ provides a convenient gravitational
master variable at the level of the equations of motion.  As shown explicitly
below, the axial system can be reduced to a closed pair of second-order master
equations for $(\Pi,\psi)$.  This field-equation reduction does not by itself
identify canonically normalized variables; that stronger question requires a
reduction of the quadratic action.}

The tangent scalar equation has already been written in covariant,
gauge-invariant form in Eq.~\eqref{CovariantOddScalarEquation}, with
$Y=\eta^2/r^2$. Here $K_Y=dK/dY$ is evaluated on the background. For
\begin{equation}
 K(Y)=\rho_0\left(\frac{Y}{Y+\mu_*^2}\right)^n,
 \label{odd_nonlinear_kinetic_function_stability}
\end{equation}
one finds
\begin{equation}
 K_Y=\rho_0 n\mu_*^2
 \frac{Y^{n-1}}{(Y+\mu_*^2)^{n+1}}.
 \label{odd_ky_stability}
\end{equation}
In the Regge--Wheeler gauge, Eq.~\eqref{CovariantOddScalarEquation}
reduces to Eq.~\eqref{ExplicitOddScalarEquation}, obtained directly in
Sec.~\ref{sec:odd_perturbations}. Together with this scalar equation, the
projected Einstein equations form the closed axial subsystem displayed in
Sec.~\ref{sec:odd_perturbations}. In particular,
{In the original projected-variable formulation, the $(tA)$ equation contains
$h_0''$ and is therefore a radial differential constraint rather than an
algebraic equation for $h_0$, while the trace-free $(AB)$ equation,
Eq.~\eqref{ExplicitOddEinsteinAB}, supplies a first-order reconstruction
relation for $\dot h_0$.  This statement refers only to the unreduced
$(h_0,h_1,\psi)$ representation.  After the gauge-invariant curl $\Pi$ is
introduced below, the $(tA)$ and $(rA)$ equations can instead be rearranged
into the algebraic reconstruction formulae
Eqs.~\eqref{Pi_h0_reconstruction}--\eqref{Pi_h1_reconstruction}.  The
quadratic-action reduction in Sec.~\ref{subsec:odd_reduced_action} then makes
the elimination of the nondynamical metric amplitudes explicit.}

The Lagrange-multiplier perturbation $\delta\lambda$ is absent from the axial
sector because it is a scalar on $S^2$.  The perturbations of $C$ and of the
three-form carry no local four-dimensional degree of freedom.  At the level
of the projected field equations they introduce no additional local axial
amplitude.  The action-level elimination of the remaining nondynamical metric
variables is carried out below in
Sec.~\ref{subsec:odd_reduced_action}.

\subsection{Gauge-invariant $\Pi$--$\psi$ master system}
\label{subsec:odd_master_equations}

{The closed axial system can be reduced directly to two field-equation master variables, represented in Regge--Wheeler gauge by $(\Pi,\psi)$.  More precisely, the gauge-invariant pair is $(\Pi,\widehat\psi)$, with $\widehat\psi=\psi$ in Regge--Wheeler gauge.  This first step does not require a kinetic normalization.  After deriving the master equations, we subsequently reduce the quadratic action in Sec.~\ref{subsec:odd_reduced_action}, which fixes a convenient local kinetic normalization and the no-ghost and radial-gradient conditions.}

{In Regge--Wheeler gauge, the gauge-invariant curl defined in Eq.~\eqref{odd_gauge_invariant_pi_stability} is, up to the overall orientation convention for $\epsilon^{ab}$,}
\begin{equation}
 {\Pi=\frac{1}{r^4}\left(\dot h_1-h_0'+\frac{2h_0}{r}\right).}
 \label{Pi_RW_explicit}
\end{equation}
{It is convenient to introduce}
\begin{equation}
 {P\equiv r^4\Pi,\qquad
 \mathcal{C}(r)\equiv16\pi G\eta^2K_Y,\qquad
 \mathcal{H}(r)\equiv\lambda_\ell+\mathcal{C}(r).}
 \label{Pi_master_definitions}
\end{equation}
{Using the background identity following from Eq.~(91),}
\begin{equation}
 {A''+\frac{2A'}{r}=\frac{\mathcal{C}}{r^2}-16\pi G K,}
 \label{background_identity_master}
\end{equation}
{the $(tA)$ and $(rA)$ Einstein equations, Eqs.~\eqref{ExplicitOddEinsteinTA} and \eqref{ExplicitOddEinsteinRA}, become}
\begin{align}
 {A\left(P'+\frac{2P}{r}\right)+\frac{\mathcal{H}}{r^2}h_0}
 &{=\mathcal{C}\dot\psi,}
 \label{Pi_algebraic_h0}\\
 {\frac{\dot P}{A}+\frac{\mathcal{H}}{r^2}h_1}
 &{=\mathcal{C}\psi'.}
 \label{Pi_algebraic_h1}
\end{align}
{Thus, wherever $\mathcal{H}\neq0$, both Regge--Wheeler metric amplitudes are reconstructed algebraically from $P$ and $\psi$:}
\begin{align}
 {h_0}
 &{=\frac{r^2}{\mathcal{H}}
 \left[\mathcal{C}\dot\psi-A\left(P'+\frac{2P}{r}\right)\right],}
 \label{Pi_h0_reconstruction}\\
 {h_1}
 &{=\frac{r^2}{\mathcal{H}}
 \left[\mathcal{C}\psi'-\frac{\dot P}{A}\right].}
 \label{Pi_h1_reconstruction}
\end{align}

{Substitution of Eqs.~\eqref{Pi_h0_reconstruction} and \eqref{Pi_h1_reconstruction} into the defining relation \eqref{Pi_RW_explicit} yields the gravitational master equation}
\begin{equation}
{
 \frac{\ddot P}{A}
 -\frac{\mathcal{H}}{r^2}\frac{d}{dr}
 \left[\frac{r^2A}{\mathcal{H}}
 \left(P'+\frac{2P}{r}\right)\right]
 +\frac{2A}{r}\left(P'+\frac{2P}{r}\right)
 +\frac{\mathcal{H}}{r^2}P
 +\frac{\lambda_\ell\mathcal{C}'}{\mathcal{H}}\dot\psi=0.}
\label{Pi_gravitational_master}
\end{equation}
{This equation is second order in the gravitational master amplitude $P=r^4\Pi$ and couples to the tangent mode through $\dot\psi$.}

{The tangent-triplet equation can be reduced by the same substitutions. Defining}
\begin{equation}
 {F(r)\equiv r^2K_Y,}
 \label{Pi_scalar_F}
\end{equation}
{Eq.~(103) becomes}
\begin{equation}
{
 -\frac{F}{A\mathcal{H}}\ddot\psi
 +\frac{d}{dr}\left(\frac{FA}{\mathcal{H}}\psi'\right)
 -K_Y\psi
 +\frac{r^2K_Y'}{\mathcal{H}^2}\dot P=0.}
\label{Pi_scalar_master}
\end{equation}
{Equations~\eqref{Pi_gravitational_master} and \eqref{Pi_scalar_master} therefore form a closed pair of coupled second-order equations for the two field-equation variables $(\Pi,\psi)$ in Regge--Wheeler gauge, with $P=r^4\Pi$.  Equivalently, before gauge fixing the gauge-invariant pair is $(\Pi,\widehat\psi)$. The metric amplitudes $h_0$ and $h_1$ are then obtained from Eqs.~\eqref{Pi_h0_reconstruction} and \eqref{Pi_h1_reconstruction}.}

{There is a useful consistency point concerning Eq.~\eqref{ExplicitOddEinsteinAB}. Because the three projected Einstein equations and the tangent-triplet equation satisfy the linearized Noether identity, Eq.~\eqref{ExplicitOddEinsteinAB} is not an additional independent evolution equation once the reduced system and the corresponding constraint are imposed. The independent master pair may therefore be obtained most transparently from the defining relation for $\Pi$ and the tangent equation after Eqs.~\eqref{ExplicitOddEinsteinTA} and \eqref{ExplicitOddEinsteinRA} have been used to reconstruct $h_0$ and $h_1$. Equation~\eqref{ExplicitOddEinsteinAB} then provides a consistency/reconstruction relation propagated by the reduced system.}

{This reduction makes the field-equation degree-of-freedom content explicit: for each radiative multipole $\ell\ge2$, the odd-parity sector is described by two coupled dynamical amplitudes, the gravitational curl $\Pi$ and the tangent scalar $\psi$.  The following subsection carries the reduction to the quadratic-action level, where a convenient local kinetic normalization and the reduced kinetic and radial-gradient matrices are obtained explicitly.}

{The derivation above provides the requested master-equation reduction at the level of the equations of motion.  We now carry the analysis one step further and derive the corresponding reduced quadratic action, which fixes a convenient local kinetic normalization and the local no-ghost and radial-gradient conditions.}

\subsection{Reduced quadratic action and local stability conditions}
\label{subsec:odd_reduced_action}

{
For the radiative multipoles $\ell\geq2$, define
\begin{equation}
 \kappa_\eta\equiv16\pi G\eta^2,
 \qquad
 F\equiv r^2K_Y,
 \qquad
 \mathcal{C}=\kappa_\eta K_Y,
 \qquad
 \mathcal{H}=\lambda_\ell+\mathcal{C}.
 \label{odd_action_definitions}
\end{equation}
In Regge--Wheeler gauge it is also convenient to introduce
\begin{equation}
 \mathcal{P}\equiv
 \dot h_1-h_0'+\frac{2h_0}{r}.
 \label{odd_action_Pscript}
\end{equation}
Before taking the second variation, the fixed-modulus constraint must also be
implemented consistently beyond first order.  Writing the tangent displacement
as $\zeta_A=\psi X_A^{\ell m}$ in Regge--Wheeler gauge, a convenient
constraint-preserving parametrization is
\begin{equation}
 \Phi^I=\eta\left[
 \left(1-\frac12\zeta_A\zeta^A\right)n^I
 +\zeta^A e_A{}^I
 \right]+O(\zeta^3),
 \label{odd_constraint_preserving_triplet}
\end{equation}
where $e_A{}^I=D_A n^I$.  Since
$n_Ie_A{}^I=0$ and $e_A{}^I e_{BI}=\Omega_{AB}$, this gives
$\Phi_I\Phi^I=\eta^2+O(\zeta^3)$.  Thus the radial second-order correction
required by the modulus constraint is retained when the matter action is
expanded.  It carries no additional odd-parity degree of freedom.

The second variation of the odd sector can be organized, up to total
$\left(t,r\right)$ derivatives, as
\begin{equation}
 S_{\mathrm{odd}}^{(2)}
 =\sum_{\ell m}\frac{L_\ell}{16\pi G}
 \int dt\,dr\;\mathcal{L}_{\mathrm{odd}}^{(2)},
 \label{odd_quadratic_action_unreduced}
\end{equation}
with
\begin{align}
 \mathcal{L}_{\mathrm{odd}}^{(2)}={}&
 \frac12\mathcal{P}^{\,2}
 +\frac{\mathcal{H}}{2Ar^2}h_0^2
 -\frac{A\mathcal{H}}{2r^2}h_1^2
 -\frac{\mathcal{C}}{A}h_0\dot\psi
 +A\mathcal{C}h_1\psi' \nonumber\\
 &+\frac{\kappa_\eta}{2}
 \left(
 \frac{F}{A}\dot\psi^{\,2}
 -FA\psi'^{\,2}
 -\lambda_\ell K_Y\psi^2
 \right).
 \label{odd_quadratic_lagrangian_unreduced}
\end{align}
The normalization of the scalar derivative terms agrees with the direct
second variation displayed below in Eq.~\eqref{odd_bare_scalar_derivative_action_stability}.
Varying Eq.~\eqref{odd_quadratic_lagrangian_unreduced} with respect to
$h_0$, $h_1$, and $\psi$, and using the background identity
Eq.~\eqref{background_identity_master}, reproduces respectively
Eqs.~\eqref{Pi_algebraic_h0}, \eqref{Pi_algebraic_h1}, and
\eqref{ExplicitOddScalarEquation}.  The trace-free angular Einstein equation
is then supplied by the linearized Noether identity, as discussed above.
Thus Eq.~\eqref{odd_quadratic_lagrangian_unreduced} is the odd quadratic
functional for the independent Regge--Wheeler-gauge amplitudes, up to boundary
terms.

To reduce the action, linearize the square in Eq.~\eqref{odd_action_Pscript}
with an auxiliary field $P$ according to
\begin{equation}
 \frac12\mathcal{P}^{\,2}
 \longrightarrow
 P\mathcal{P}-\frac12P^2.
 \label{odd_action_auxiliary_P}
\end{equation}
The $P$ equation sets $P=\mathcal{P}=r^4\Pi$.  After integrating the
$P h_0'$ and $P\dot h_1$ terms by parts, both $h_0$ and $h_1$ become
algebraic.  Their equations are precisely
Eqs.~\eqref{Pi_h0_reconstruction} and \eqref{Pi_h1_reconstruction}.
Substituting these solutions back into the action gives
\begin{align}
 \mathcal{L}_{\mathrm{red}}^{(2)}={}&
 -\frac12P^2
 +\frac{r^2}{2A\mathcal{H}}
 \left(\dot P-A\mathcal{C}\psi'\right)^2
 -\frac{Ar^2}{2\mathcal{H}}
 \left(P'+\frac{2P}{r}-\frac{\mathcal{C}}{A}\dot\psi\right)^2
 \nonumber\\
 &+\frac{\kappa_\eta}{2}
 \left(
 \frac{F}{A}\dot\psi^{\,2}
 -FA\psi'^{\,2}
 -\lambda_\ell K_Y\psi^2
 \right).
 \label{odd_reduced_quadratic_action_compact}
\end{align}
The Euler--Lagrange equations obtained from
Eq.~\eqref{odd_reduced_quadratic_action_compact} are exactly the coupled
master equations \eqref{Pi_gravitational_master} and
\eqref{Pi_scalar_master}.

Expanding the squares gives the equivalent form
\begin{align}
 \mathcal{L}_{\mathrm{red}}^{(2)}={}&
 \frac{r^2}{2A\mathcal{H}}\dot P^{\,2}
 -\frac{Ar^2}{2\mathcal{H}}
 \left(P'+\frac{2P}{r}\right)^2
 -\frac12P^2
 \nonumber\\
 &+\frac{\kappa_\eta\lambda_\ell F}{2A\mathcal{H}}\dot\psi^{\,2}
 -\frac{\kappa_\eta\lambda_\ell FA}{2\mathcal{H}}\psi'^{\,2}
 -\frac{\kappa_\eta\lambda_\ell K_Y}{2}\psi^2
 \nonumber\\
 &+\frac{r^2\mathcal{C}}{\mathcal{H}}
 \left[
 \left(P'+\frac{2P}{r}\right)\dot\psi
 -\dot P\,\psi'
 \right].
 \label{odd_reduced_quadratic_action_expanded}
\end{align}
The last line is antisymmetric in the principal derivatives.  Up to a total
derivative it may be written as
\begin{equation}
 \frac{r^2\mathcal{C}}{\mathcal{H}}
 \left[
 \left(P'+\frac{2P}{r}\right)\dot\psi-\dot P\psi'
 \right]
 \doteq
 -\frac{\lambda_\ell r^2\mathcal{C}'}{\mathcal{H}^2}
 P\dot\psi,
 \label{odd_gyroscopic_term}
\end{equation}
where $\doteq$ denotes equality up to boundary terms.  Hence this mixing does
not alter the symmetric principal kinetic or radial-gradient matrices.

Introducing the tortoise coordinate $dr_*/dr=A^{-1}$, the principal part of
the reduced action is
\begin{equation}
 S_{\mathrm{odd,prin}}^{(2)}
 =\sum_{\ell m}\frac{L_\ell}{32\pi G}
 \int dt\,dr_*
 \left[
 \frac{r^2}{\mathcal{H}}
 \left(\dot P^{\,2}-P_{,*}^{\,2}\right)
 +\frac{\kappa_\eta\lambda_\ell r^2K_Y}{\mathcal{H}}
 \left(\dot\psi^{\,2}-\psi_{,*}^{\,2}\right)
 \right]
 +\hbox{lower-derivative terms}.
 \label{odd_reduced_action_principal}
\end{equation}
The kinetic and radial-gradient matrices are therefore diagonal and have the
same positive eigenvalues, apart from the common positive harmonic factor,
provided
\begin{equation}
 \mathcal{H}>0,
 \qquad
 K_Y>0.
 \label{odd_noghost_conditions}
\end{equation}
For $\ell\geq2$ one has $\lambda_\ell>0$.  In the regular black-hole branch
considered here $K_Y>0$, so that
$\mathcal{H}=\lambda_\ell+\kappa_\eta K_Y>0$ automatically.  Consequently,
at every finite radius in the nonextremal exterior, both reduced kinetic
eigenvalues are positive and the two radial characteristic speeds are
\begin{equation}
 c_{r,g}^2=c_{r,s}^2=1.
 \label{odd_radial_speeds}
\end{equation}
A convenient local kinetic normalization is
\begin{equation}
 Q_g=\sqrt{\frac{L_\ell}{16\pi G}}\,
 \frac{r}{\sqrt{\mathcal{H}}}\,P,
 \qquad
 Q_s=\eta\sqrt{L_\ell\lambda_\ell}\,
 \frac{r\sqrt{K_Y}}{\sqrt{\mathcal{H}}}\,\psi.
 \label{odd_canonical_variables}
\end{equation}
Because the normalization factors depend only on $r$, these variables put the
diagonal quadratic time-derivative terms in unit-normalized form; their radial
dependence is absorbed into lower-derivative potential and mixing terms.  We
refer to this as a local kinetic normalization rather than a complete canonical
transformation: the first-order gyroscopic coupling in
Eq.~\eqref{odd_gyroscopic_term} remains and therefore also enters the
conjugate momenta.

There is nevertheless an important asymptotic qualification.  For the
$n=3$ solution, Eq.~\eqref{odd_ky_asymptotic_stability} gives
$K_Y=O(r^{-4})$, so the scalar kinetic eigenvalue in
Eq.~\eqref{odd_reduced_action_principal} behaves as $r^2K_Y=O(r^{-2})$ and
tends to zero as $r\to\infty$.  Thus the reduced action is positive at every
finite exterior radius but is not uniformly nondegenerate on the complete
asymptotic domain.  The quadratic action therefore establishes the local
absence of ghosts and radial-gradient instabilities in the finite exterior,
but it does not by itself prove mode stability or exclude an asymptotic
strong-coupling issue.  Those stronger questions require analysis of the
full coupled potential operator and its boundary conditions.
}

\subsection{Constraint structure and dynamical variables}
\label{subsec:odd_constraint_structure}

The equations distinguish three notions that should not be conflated.  A field
with no independent second-order time derivative is nondynamical in the
corresponding field-equation system.  A Lagrange multiplier is a variable that
appears linearly and without derivatives in an action and imposes a constraint.
An auxiliary field is eliminated algebraically by its Euler--Lagrange equation.
Finally, a radial differential constraint must be solved subject to radial
boundary data and is not an algebraic elimination.

{With these definitions, the original projected equations contain a radial
differential constraint and a first-order reconstruction relation.  After
introducing the gauge-invariant curl $\Pi$, however,
Eqs.~\eqref{ExplicitOddEinsteinTA} and \eqref{ExplicitOddEinsteinRA} can be
rearranged into the algebraic reconstruction formulae
\eqref{Pi_h0_reconstruction} and \eqref{Pi_h1_reconstruction}.  The radiative
odd-parity dynamics is therefore represented at the field-equation level by
the closed master pair $(\Pi,\psi)$ derived above.  This result should not be
confused with the action-level step itself.  The reduced quadratic action
derived in Sec.~\ref{subsec:odd_reduced_action} shows that the two kinetic and
radial-gradient eigenvalues are positive at every finite exterior radius on
the regular branch with $K_Y>0$, while the scalar eigenvalue tends to zero
asymptotically.}

The field equations nevertheless establish that the tangent mode has a
second-order evolution equation.  Retaining only the terms with two
derivatives of $\psi$ in Eq.~\eqref{ExplicitOddScalarEquation} gives
\begin{equation}
 \begin{split}
 -\frac{r^2K_Y}{A}\,\ddot\psi
 +r^2 A K_Y\,\psi''
 +{}&\text{terms with fewer derivatives of $\psi$}\\
 &\text{and metric couplings}=0.
 \end{split}
 \label{odd_scalar_principal_part_stability}
\end{equation}
For $K_Y\neq0$, its characteristic cone coincides with the radial null cone of
the background metric.  This is a statement about the unreduced scalar
equation.  The reduced action in Sec.~\ref{subsec:odd_reduced_action} shows explicitly
that the symmetric principal kinetic and radial-gradient matrices are diagonal
and positive for $K_Y>0$ and $\mathcal H>0$.  The derivative mixing is
antisymmetric in the principal derivatives and reduces, up to boundary terms,
to the lower-derivative term in Eq.~\eqref{odd_gyroscopic_term}.

\subsection{Bare scalar kinetic term and its limitations}
\label{subsec:odd_bare_scalar_kinetic}

The part of the matter action quadratic in derivatives of the tangent
perturbation follows directly from the second variation of $Y$.  With the
harmonic normalization
\begin{equation}
 \int_{S^2}d\Omega\,
 X_A^{\ell m}X^{A\ell m*}=L_\ell,
 \label{odd_vector_harmonic_normalization_stability}
\end{equation}
it is
\begin{equation}
 S_{\psi,\mathrm{der}}^{(2)}
 =\frac{\eta^2L_\ell}{2}
 \int dt\,dr\,r^2K_Y
 \left(\frac{\dot\psi^{\,2}}{A}-A\psi'^{,2}\right),
 \label{odd_bare_scalar_derivative_action_stability}
\end{equation}
up to terms involving the metric perturbations and terms without derivatives
of $\psi$.  Equation~\eqref{odd_bare_scalar_derivative_action_stability} is
not the reduced matter--gravity action.  It establishes only the sign and the
radial characteristic speed of the bare tangent-scalar term.

For $\rho_0>0$, $n>0$, and $Y>0$,
Eq.~\eqref{odd_ky_stability} gives $K_Y>0$.  Hence the bare scalar time-kinetic
term in the black-hole exterior, where $A>0$, has the conventional sign, and
its bare radial gradient term has the corresponding hyperbolic sign.  By itself, this bare term would not prove the absence of ghosts or gradient
instabilities in the full linearized system.  The reduced quadratic action in
Sec.~\ref{subsec:odd_reduced_action} supplies the constraint reduction and
shows that, at every finite exterior radius, the two symmetric kinetic and
radial-gradient eigenvalues are positive on the regular branch with
$K_Y>0$.

For the $n=3$ background used in Sec.~\ref{sec:background},
\begin{equation}
 K_Y=3\rho_0\mu_*^2
 \frac{Y^2}{(Y+\mu_*^2)^4}.
 \label{odd_ky_n3_stability}
\end{equation}
At a regular nonextremal event horizon $r=r_+$, $Y$ and $K_Y$ are finite.
Introducing $dr_*/dr=A^{-1}$, the derivative part of the scalar action becomes
\begin{equation}
 S_{\psi,\mathrm{der}}^{(2)}
 =\frac{\eta^2L_\ell}{2}
 \int dt\,dr_*\,r^2K_Y
 \left[(\partial_t\psi)^2-(\partial_{r_*}\psi)^2\right],
 \label{odd_bare_scalar_tortoise_action_stability}
\end{equation}
so its principal coefficient is regular at $r_+$ in tortoise coordinates.
At large radius,
\begin{equation}
 Y=\frac{\eta^2}{r^2},
 \qquad
 K_Y=\frac{3\rho_0\eta^4}{\mu_*^6r^4}
 +O(r^{-6}),
 \qquad r\rightarrow\infty,
 \label{odd_ky_asymptotic_stability}
\end{equation}
and therefore $r^2K_Y=O(r^{-2})$.  The bare tangent-scalar normalization thus
vanishes asymptotically.  Whether a canonically normalized scalar variable is
regular there, or whether this instead signals a genuinely degenerate or
strongly coupled asymptotic sector, cannot be decided from the unreduced term
alone.  This behavior agrees with the reduced-action result: the scalar kinetic
eigenvalue remains positive at each finite exterior radius but approaches zero
at infinity, so the kinetic operator is not uniformly nondegenerate on the
complete asymptotic domain.
\subsection{Radial behavior of the direct metric--scalar mixing coefficient}
\label{subsec:axial_coupling_localization}

The coupling between the axial metric perturbations and the tangent
scalar mode can be characterized directly from the projected field equations.
The reduced quadratic action derived in Sec.~\ref{subsec:odd_reduced_action}
provides the corresponding canonical interpretation; here we retain the
original field-equation coefficient as a useful radial diagnostic. For the
$n=3$ background, the nonlinear matter function evaluated on the
hedgehog solution is
\begin{equation}
K(r)
=
\rho_0
\left(
\frac{L^2}{r^2+L^2}
\right)^3.
\label{n3_matter_profile}
\end{equation}
Introducing the dimensionless radius
\begin{equation}
x=\frac{r}{L},
\end{equation}
this expression becomes
\begin{equation}
\frac{K(x)}{\rho_0}
=
\frac{1}{(1+x^2)^3}.
\label{dimensionless_matter_profile}
\end{equation}
The matter density is therefore concentrated in the regular-core
region and decreases as $x^{-6}$ at large radius.

The direct metric--scalar mixing in the projected axial Einstein
equations is proportional to $\eta^2K_Y$. For
\begin{equation}
K(Y)
=
\rho_0
\left(
\frac{Y}{Y+\mu_*^2}
\right)^3,
\qquad
Y=\frac{\eta^2}{r^2},
\qquad
L=\frac{\eta}{\mu_*},
\end{equation}
one obtains
\begin{equation}
\eta^2K_Y
=
3\rho_0L^2
\frac{x^4}{(1+x^2)^4}.
\label{eta2KY_radial_profile}
\end{equation}
It is useful to define the dimensionless direct axial mixing coefficient
\begin{equation}
\mathcal{C}_{\mathrm{ax}}(x)
\equiv
16\pi G\eta^2K_Y.
\label{axial_mixing_definition}
\end{equation}
Using
\begin{equation}
M
=
\frac{\pi^2}{4}\rho_0L^3,
\qquad
\chi=\frac{GM}{L},
\end{equation}
Eq.~\eqref{axial_mixing_definition} becomes
\begin{equation}
\mathcal{C}_{\mathrm{ax}}(x)
=
\frac{192\chi}{\pi}
\frac{x^4}{(1+x^2)^4}.
\label{axial_mixing_profile}
\end{equation}

The function $x^4/(1+x^2)^4$ vanishes as $x^4$ near the
center, reaches its maximum at $x=1$, and decreases monotonically for
$x>1$. The outer horizon of the black-hole branch satisfies
\begin{equation}\label{209}
x_+\geq x_{\mathrm{ext}}
\simeq 1.18924>1.
\end{equation}
Consequently, throughout the exterior region $x\geq x_+$,
the coefficient $\mathcal{C}_{\mathrm{ax}}(x)$ attains its largest
value at the event horizon and decreases monotonically with radius.
This statement concerns the coefficient multiplying the direct
metric--scalar mixing terms in the projected field equations. It
should not be interpreted as a statement about a canonically
normalized interaction strength, which can be determined only after
reducing the complete quadratic action. Its asymptotic behavior is
\begin{equation}
\mathcal{C}_{\mathrm{ax}}(x)
=
\frac{192\chi}{\pi x^4}
+
\mathcal{O}(x^{-6}),
\qquad
x\rightarrow\infty.
\label{axial_mixing_asymptotic}
\end{equation}

The direct metric--scalar mixing coefficient is therefore largest at the
event horizon and becomes rapidly suppressed far from the black hole.
This radial behavior refers only to the coefficient in the projected
field equations; it does not establish localization of the complete
physical interaction after canonical reduction and does not by itself measure
the strength of the interaction between canonically normalized modes. It also explains why the
vacuum Regge--Wheeler structure is recovered asymptotically, although
the metric and scalar perturbations remain coupled at finite radius.

This radial behavior alone does not determine a quasinormal-mode
frequency shift or prove stability; those questions require the
reduced coupled operator and its boundary conditions.
\subsection{Schwarzschild consistency check}
\label{subsec:odd_schwarzschild_check}

The vacuum limit requires both the background matter stress tensor and the
metric--scalar mixing to vanish.  Thus one takes
\begin{equation}
 K\rightarrow0,
 \qquad
 \eta^2K_Y\rightarrow0,
 \qquad
 A(r)\rightarrow1-\frac{2GM}{r}.
 \label{odd_vacuum_limit_stability}
\end{equation}
The projected Einstein equations of Sec.~\ref{sec:odd_perturbations} then
reduce to the standard axial vacuum equations. The scalar equation,
however, simultaneously loses its overall kinetic coefficient as
$K_Y\to0$, so the full coupled operator changes rank in this limit; only
the metric block has a smooth Regge--Wheeler reduction. Eliminating the constraint in
that limit and defining the
Regge--Wheeler variable
\begin{equation}
 \Psi_{\mathrm{RW}}=\frac{A h_1}{r}
 \label{odd_rw_variable_stability}
\end{equation}
gives
\begin{equation}
 \left[-\partial_t^2+\partial_{r_*}^2
 -V_{\mathrm{RW}}(r)\right]\Psi_{\mathrm{RW}}=0,
 \label{odd_rw_equation_stability}
\end{equation}
with
\begin{equation}
 V_{\mathrm{RW}}(r)
 =\left(1-\frac{2GM}{r}\right)
 \left[\frac{L_\ell}{r^2}-\frac{6GM}{r^3}\right].
 \label{odd_rw_potential_stability}
\end{equation}
This potential vanishes at the Schwarzschild horizon and behaves as
$L_\ell/r^2+O(r^{-3})$ at infinity.  Equations~\eqref{odd_rw_equation_stability} and
\eqref{odd_rw_potential_stability} check the vacuum normalization of
the explicitly displayed metric equations.  They do not determine the potential matrix of the coupled
regular-black-hole system, because the scalar mixing survives away
from the vacuum limit~\eqref{odd_vacuum_limit_stability}.

\subsection{Scope of the stability statement}
\label{subsec:odd_stability_scope}

The closed axial equations support the following limited conclusions. They can be
written in terms of the gravitational curl $\Pi$ and the tangent amplitude
$\widehat\psi$, subject to the projected Einstein constraint. The projected
tangent equation is second order and has
the background radial null cone wherever $K_Y\neq0$.  For the parameter range
$\rho_0>0$, $n>0$, and $Y>0$, the bare tangent-scalar kinetic term has the
conventional sign.  The metric equations possess the correct
Regge--Wheeler limit when the matter sector is removed.

Parity decoupling follows from the simultaneous polar--axial calculation,
and the reduced quadratic action now establishes the local no-ghost and
radial-gradient conditions for the two odd-parity modes.  In particular, for
$K_Y>0$ and $\ell\ge2$ the two reduced kinetic eigenvalues and the two radial-gradient eigenvalues are positive at every finite exterior radius, with unit
radial characteristic speeds.  This does not constitute a complete mode-stability
theorem: the scalar kinetic eigenvalue tends to zero asymptotically, and we
have not established positivity of the full coupled potential operator or the
absence of exponentially growing modes under the physical boundary conditions.
\section{Discussion and Conclusions}
\label{sec:conclusions}

In this study, we investigated linear perturbations of a geometrically
regular black hole sourced by a constrained $\mathrm{SO}(3)$ scalar
triplet and an auxiliary three-form field. The main difficulty is that
the individual background scalars depend explicitly on $(\theta,\phi)$,
although the metric and the total energy--momentum tensor are spherically
symmetric. For this reason, the separation between polar and axial
perturbations was not assumed in advance. Both sectors were introduced
simultaneously and their projections were calculated from the same
unrestricted perturbative ansatz. The main results can be summarized as
follows:

\begin{itemize}

\item The fixed-modulus constraint eliminates the perturbation normal to
the internal two-sphere. The remaining tangent perturbation separates
into a polar amplitude and an axial amplitude. The Lagrange multiplier
and three-form perturbations supply no additional local axial radiative
mode.

\item Direct variation of the kinetic invariant gives
$\delta Y_{\mathrm{ax}}=0$. Consequently, the variations of $K$ and
$K_Y$ contain no axial contribution. The polar energy--momentum
projections contain only polar harmonics, whereas the axial projections
contain only axial harmonics.

\item The tangent-triplet field equation preserves the gradient and
axial-vector harmonic subspaces separately. Together with the
corresponding property of the linearized Einstein tensor, this makes the
two off-diagonal blocks of the radiative Einstein--triplet operator vanish
for $\ell\geq2$. Therefore, polar and axial radiative perturbations decouple
at linear order for the action considered here, while the three-form
constraint sector introduces no independent local radiative amplitude.

{\item For each radiative multipole $\ell\geq2$, the axial system can be
reduced explicitly to a closed pair of coupled second-order master equations
for the gauge-invariant gravitational curl $\Pi$ and the tangent scalar
amplitude $\psi$.  The Regge--Wheeler metric amplitudes $h_0$ and $h_1$ are
reconstructed algebraically from these master variables through
Eqs.~\eqref{Pi_h0_reconstruction} and \eqref{Pi_h1_reconstruction}.
Thus, at the level of the equations of motion, the odd-parity radiative sector
contains two coupled dynamical amplitudes.}

\item The tangent-triplet equation has a second-order hyperbolic
principal part wherever $K_Y\neq0$. For the nonlinear kinetic function
used in the regular solution, $\rho_0>0$, $n>0$, and $Y>0$ imply
$K_Y>0$. Hence, the bare scalar time-derivative term has the conventional
sign in the black-hole exterior, and its radial characteristic cone
coincides with the radial null cone of the background.

\item For the $n=3$ solution, the coefficient multiplying the direct
axial metric--scalar mixing terms is
\[
\mathcal C_{\mathrm{ax}}(x)
=\frac{192\chi}{\pi}\frac{x^4}{(1+x^2)^4}.
\]
Since $x_+\geq x_{\mathrm{ext}}>1$, this coefficient has its largest
exterior value at the event horizon and decreases monotonically
outward, falling as $r^{-4}$ at large distance. This result concerns
the direct metric--scalar mixing coefficient in the projected field
equations; it does not establish localization of the canonically reduced
physical interaction.

\item When the background matter contribution and the metric--scalar
mixing are removed, the axial metric block reduces to the standard
Regge--Wheeler system. The scalar kinetic coefficient vanishes in the
same limit, so the full coupled operator changes rank rather than reducing
smoothly. The resulting metric wave equation and potential,
Eqs.~\eqref{odd_rw_equation_stability} and
\eqref{odd_rw_potential_stability}, confirm the normalization and the
GR limit of the projected metric equations.

\end{itemize}

{The above results establish linear parity decoupling and, in addition,
reduce the radiative odd-parity sector explicitly to a closed pair of
coupled second-order master equations for $(\Pi,\psi)$.  In the original
$(h_0,h_1,\psi)$ formulation the projected Einstein equations contain
constraint and reconstruction relations; after introducing $P=r^4\Pi$,
however, $h_0$ and $h_1$ are reconstructed algebraically from the master
variables through Eqs.~\eqref{Pi_h0_reconstruction} and
\eqref{Pi_h1_reconstruction}.  This establishes the dynamical field-equation content of the odd sector.
The reduced quadratic action then shows that, on the regular branch with
$K_Y>0$, both kinetic and radial-gradient eigenvalues are positive at every
finite exterior radius and provides a convenient local kinetic normalization.  The
scalar kinetic eigenvalue nevertheless tends to zero asymptotically, and the
sign of the full coupled potential operator has not been established; these
facts prevent us from claiming a complete mode-stability theorem.}

The quadratic-action reduction carried out here fixes the two local kinetic
and radial-gradient eigenvalues and identifies a convenient local kinetic
normalization for the odd sector.  A stronger global stability analysis still
requires the full coupled lower-derivative potential operator and the physical
boundary conditions, particularly in view of the asymptotic degeneration of
the scalar normalization.  Establishing positivity of that operator, excluding
exponentially growing modes, and computing a reliable quasinormal-mode spectrum
are therefore left for a separate analysis.

Finally, the present dynamical discussion is restricted to radiative
odd-parity perturbations with $\ell\geq2$ in the black-hole exterior. The
dipole sector, the dynamical reduction of the polar equations, the
interior and Cauchy-horizon problems, and the extremal background remain
open questions. These problems provide natural directions in which the
present investigation can be extended.
\section*{Acknowledgments}

The author would like to thank Sebastian Bahamonde for valuable discussions, careful comments, and constructive suggestions that significantly improved the content and presentation of this work.
\newpage
\appendix
\section{Projection of the Odd-Parity Field Equations}
\label{AppOddReduction}

In this appendix we summarize the projection of the linearized
Einstein--matter equations onto the axial vector and tensor spherical
harmonics. The explicit Regge--Wheeler-gauge component equations are
given in Sec.~\ref{sec:odd_perturbations}. Here we record the projection
conventions and the
covariant origin of those equations. No action-level reduction is
performed. Throughout, all background quantities are evaluated on the
static solution introduced in Sec.~\ref{sec:background}.

For each mode labelled by $(\ell,m)$, the odd-parity perturbations take
the form
\begin{align}
h_{aA}
&=
h_a(t,r)X_A^{\ell m},
\qquad
h_{AB}
=
h_2(t,r)X_{AB}^{\ell m},\nonumber\\
\delta\Phi^I
&=
\eta\,\psi(t,r)X^{A,\ell m}e_A^{I}.
\end{align}
The Regge--Wheeler gauge is imposed only after the projected equations
have been obtained.

The variation of the Levi--Civita connection is
\begin{equation}
\delta\Gamma^{\rho}{}_{\mu\nu}
=
\frac{1}{2}\bar g^{\rho\sigma}
\left(
\bar\nabla_{\mu}h_{\nu\sigma}
+
\bar\nabla_{\nu}h_{\mu\sigma}
-
\bar\nabla_{\sigma}h_{\mu\nu}
\right).
\label{AppDeltaGamma}
\end{equation}
and the corresponding variation of the Ricci tensor is
\begin{equation}
\delta R_{\mu\nu}
=
\frac{1}{2}
\left(
\bar\nabla_{\rho}\bar\nabla_{\mu}h^{\rho}{}_{\nu}
+
\bar\nabla_{\rho}\bar\nabla_{\nu}h^{\rho}{}_{\mu}
-
\bar\Box h_{\mu\nu}
-
\bar\nabla_{\mu}\bar\nabla_{\nu}h
\right).
\label{AppDeltaRicci}
\end{equation}
At first order, the Einstein tensor reads
\begin{equation}
\delta G_{\mu\nu}
=
\delta R_{\mu\nu}
-
\frac{1}{2}\bar{g}_{\mu\nu}\,\delta R
-
\frac{1}{2}\bar{R}\,h_{\mu\nu},
\label{AppDeltaEinstein}
\end{equation}
and the perturbation of the Ricci scalar is
\begin{equation}
\delta R
=
\bar\nabla_{\mu}\bar\nabla_{\nu}h^{\mu\nu}
-
\bar\Box h
-
h^{\mu\nu}\bar R_{\mu\nu}.
\label{AppDeltaRicciScalar}
\end{equation}

For odd parity, we need the $(tA)$ and $(rA)$ equations, together with
the trace-free $(AB)$ equation. Starting from
\begin{equation}
\delta G_{\mu\nu}
=
8\pi G\,\delta T_{\mu\nu}.
\label{AppLinearizedEinstein}
\end{equation}
For each component, the harmonic coefficients are defined by
\begin{align}
\mathcal{G}_{a}
&=
\frac{1}{L_\ell}
\int_{S^2}d\Omega\,
X^{A\,\ell m *}\delta G_{aA},
&
\mathcal{T}_{a}
&=
\frac{1}{L_\ell}
\int_{S^2}d\Omega\,
X^{A\,\ell m *}\delta T_{aA},
\\
\mathcal{G}_{2}
&=
\frac{2}{L_\ell\lambda_\ell}
\int_{S^2}d\Omega\,
X^{AB\,\ell m *}\delta G_{AB}^{\mathrm{TF}},\nonumber\\
&
\mathcal{T}_{2}
=
\frac{2}{L_\ell\lambda_\ell}
\int_{S^2}d\Omega\,
X^{AB\,\ell m *}\delta T_{AB}^{\mathrm{TF}}.
\end{align}
Here,
\begin{equation}
L_\ell=\ell(\ell+1),
\qquad
\lambda_\ell=(\ell-1)(\ell+2).
\end{equation}

After the angular projections are performed, the remaining equations
involve only $t$ and $r$. We denote them by
\begin{align}
\mathcal{G}_{t}[h_0,h_1,h_2]
&=
8\pi G\,
\mathcal{T}_{t}[h_0,h_1,h_2,\psi],
\label{AppOddEt}
\\
\mathcal{G}_{r}[h_0,h_1,h_2]
&=
8\pi G\,
\mathcal{T}_{r}[h_0,h_1,h_2,\psi],
\label{AppOddEr}
\\
\mathcal{G}_{2}[h_0,h_1,h_2]
&=
8\pi G\,
\mathcal{T}_{2}[h_0,h_1,h_2,\psi].
\label{AppOddE2}
\end{align}

Here the symbols
$\mathcal{G}_{t}$,
$\mathcal{G}_{r}$,
$\mathcal{G}_{2}$,
and
$\mathcal{T}_{t}$,
$\mathcal{T}_{r}$,
$\mathcal{T}_{2}$
denote the differential expressions obtained from the projected $(tA)$,
$(rA)$, and trace-free $(AB)$ components. They are shorthand for the
explicit equations displayed in
Eqs.~\eqref{ExplicitOddEinsteinTA}--\eqref{ExplicitOddEinsteinAB}; the
schematic notation in
Eqs.~\eqref{AppOddEt}--\eqref{AppOddE2} does not constitute an
additional reduction or a set of master equations.

The tangent scalar equation is obtained independently from the
variation of the scalar action. In the Regge--Wheeler gauge, it is
given by Eq.~\eqref{ExplicitOddScalarEquation} of the main text.

{In the unreduced $(h_0,h_1,\psi)$ variables, the projected $(tA)$ equation
contains radial derivatives of $h_0$, and the trace-free $(AB)$ equation
supplies the reconstruction relation for $\dot h_0$ given in
Eq.~\eqref{ExplicitOddEinsteinAB}.  This does not prevent a field-equation
master reduction: after introducing $P=r^4\Pi$, the $(tA)$ and $(rA)$
projections yield the algebraic reconstruction formulae
Eqs.~\eqref{Pi_h0_reconstruction} and \eqref{Pi_h1_reconstruction}, and the
radiative sector is governed by the closed master pair
Eqs.~\eqref{Pi_gravitational_master} and \eqref{Pi_scalar_master}.  The projected equations alone do not supply the action-level elimination of
nondynamical variables.  That additional step is performed explicitly in
Sec.~\ref{subsec:odd_reduced_action}, where the reduced kinetic and gradient
matrices and a convenient local kinetic normalization are obtained.}

%

\end{document}